# Educational Data mining and Learning Analytics: An updated survey

**First author: Full name and affiliation; plus email address if corresponding author**
Cristobal Romero
University of Cordoba
cromero@uco.es

**Second author: Full name and affiliation; plus email address if corresponding author**
Sebastian Ventura
University of Cordoba

## KeyWords:

## Abstract

This survey is an updated and improved version of the previous one published in 2013 in this journal with the title Data Mining in Education. It reviews in a comprehensible and very general way how Educational Data Mining and Learning Analytics have been applied over educational data. In the last decade, this research area has evolved enormously and a wide range of related terms are now used in the bibliography such as Academic Analytics, Institutional Analytics, Teaching Analytics, Data-Driven Education, Data-Driven Decision Making in Education, Big Data in Education and Educational Data Science. This paper provides the current state of the art by reviewing the main publications, the key milestones, the knowledge discovery cycle, the main educational environments, the specific tools, the free available datasets, the most used methods, the main objectives and the future trends in this research area.

## INTRODUCTION

The increase of e-learning resources, instrumental educational software, the use of the Internet in education, and the establishment of state databases of student information has created large repositories of educational data. Traditional Educational institutions have used for many year information systems that store plenty of interesting information. Nowadays, Web-based educational systems have been rising exponentially and they led us to store a huge amount of potential data from multiple sources with different formats and with different granularity levels (Romero and Ventura, 2017). And new types of educational environments such as Blended Learning, Virtual/Enhanced environments, Mobile/Ubiquitous learning, Game Learning, etc. also gather huge amount of data about students. All these systems produce huge amount of information of high educational value, but it is impossible to analyze it manually. So, tools to automatically analyze this kind of data are needed because of all this information provides a goldmine of educational data that can be explored and exploited to understand how students learn. In fact, today, one of the biggest challenges that educational institutions face is the exponential growth of educational data and the transformation of this data into new insights that can benefit students, teachers, and administrators (Baker, 2015).

Two different communities have grown around the same area with a joint interest in how educational data can be exploited to benefit education and the science of learning (Baker and Inventado, 2014):

- **Educational Data Mining (EDM)** is concerned with developing methods for exploring the unique types of data that come from educational environments (Bakhshinategh et al. 2018). It can be also defined as the application of Data Mining (DM) techniques to this specific type of dataset that come from educational environments to address important educational questions (Romero and Ventura, 2013).

- **Learning Analytics (LA)** can be defined as the measurement, collection, analysis and reporting of data about learners and their contexts, for purposes of understanding and optimising learning and the environments in which it occurs (Lang et al., 2017). There are three crucial elements involved in this definition (Siemens, 2013): data, analysis and action.

Both communities share a common interest in data-intensive approaches to educational research, and share the goal of enhancing educational practice (Siemens & Baker, 2012) (Liñan et al. 2015). On the one hand, LA is focused on the educational challenge and EDM is focused on the technological challenge. LA is focused on data-driven decision making and integrating the technical and the social/pedagogical dimensions of learning by applying known predictive models. On the other hand, EDM is generally looking for new patterns in data and developing new algorithms and/or models. Ultimately, the differences between the two communities are more based on focus, research questions, and the eventual use of models, than on the methods used (Baker and Inventado, 2014). Regardless of the differences between the LA and EDM communities, the two areas have significant overlap both in the objectives of investigators as well as in the methods and techniques that are used in the investigation.

EDM and LA are interdisciplinary areas including but not limited to information retrieval, recommender systems, visual data analytics, domain-driven data mining, social network analysis, psychopedagogy, cognitive psychology, psychometrics, etc. In fact, they can be drawn as the combination of three main areas (see Figure 1): computer science, education and statistics. The intersection of these three areas also forms other subareas closely related to EDM and LA such as computer-based education, data mining & machine learning, and educational statistics.

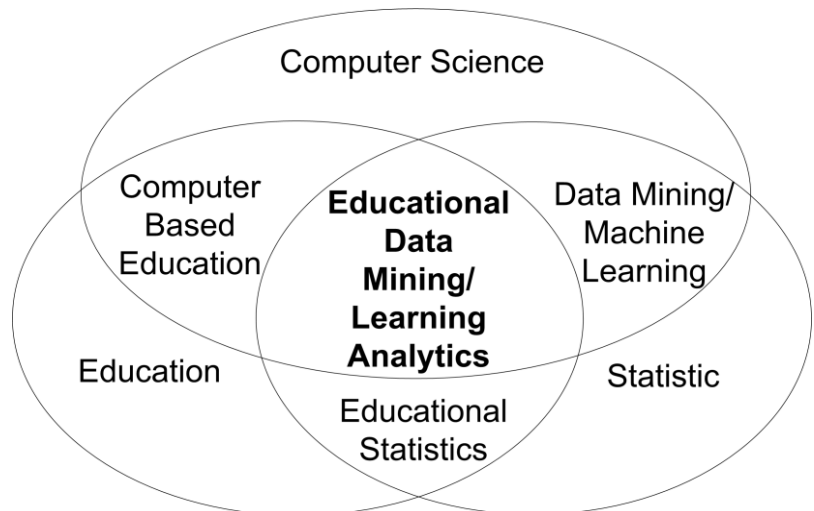

Figure 1. Main areas related to EDM/LA.

In addition to EDM and LA, there are also other related terms used in the bibliography:

- **Academic Analytics (AA) and Institutional Analytics (IA)** is concerned with the collection, analysis, and visualization of academic program activities such as courses, degree programs; research, revenue of students' fees, course evaluation, resource allocation and management to generate institutional insight (Campbell et al. 2007) (Long and Siemens, 2014). So, it is focused on the political/economic challenge.

- **Teaching Analytics (TA)** refers to the analysis of teaching activities and performance data as well as the design, development and evaluation of teaching activities (Prieto et al. 2016). It is focused on the educational challenge from the instructors' point of view.

- **Data-Driven Education (DDE) and Data-Driven Decision Making in Education (DDDM)** refers to systematically collect and analyze various types of educational data, to guide a range of decisions to help improve the success of students and schools (Custer et al. 2018)(Datnow and Hubbard, 2016).

- **Big Data in Education (BDE)** refers to apply big data (basic connotation summed up in volume, variety, value and velocity) techniques to data from educational environment (Daniel, 2019).

- **Educational Data Science (EDS)** is defined as the use of data gathered from educational environments/settings for solving educational problems (Romero & Ventura, 2017). Data science is a concept to unify statistics, data analysis, machine learning and their related methods.

This survey is an updated and improved version of the previous one published in 2013 in this journal with the title Data Mining in Education (Romero and Ventura, 2013). It is needed to redo a comprehensible overview of the current state of knowledge in EDM and LA because six years have passed and a huge number of new papers have been published. The main changes we have observed since the previous survey are: new related terms are used in the bibliography (AA, IA, TA, DDE, DDDM, BDE and EDS), the number of published books and papers has grown exponentially (more in

LA than EDM), the interest on data related to new types of educational environments has increased (MOOCs, virtual and augmented reality learning, serious games, blended learning, etc.), more specific tools and free datasets are available, the number of application problems or topics of interest is wider, and finally, there are new future trends.

## BACKGROUND

EDM and LA have emerged from two independent conferences and communities. The first Educational Data Mining conference was in Montreal, Canada in 2008 organized by the IEDM society, and the first Learning Analytics and Knowledge conference was in Banff, Canada in 2011 organized by the SOLAR society. There are also some other closely-related conferences (see Table 1).

| Title | Acronym | Type | 1º Year |
|---|---|---|---|
| International Conference on Artificial Intelligence in Education | AIED | Bi-Annual | 1982 |
| International Conference on Intelligent Tutoring Systems | ITS | Bi-Annual | 1988 |
| IEEE International Conference on Advanced Learning Technologies | ICALT | Annual | 2000 |
| European Conference on Technology-Enhanced Learning | EC-TEL | Annual | 2006 |
| International Conference on Educational Data Mining | EDM | Annual | 2008 |
| International Conference on User Modeling, Adaptation, and Personalization | UMAP | Annual | 2009 |
| International Conference on Learning Analytics and Knowledge | LAK | Annual | 2011 |
| Learning at Scale | L@S | Annual | 2014 |
| Learning and Students Analytics Conference | LSAC | Annual | 2017 |

Table 1. Most related conferences about EDM/LA.

The first book about EDM/LA topics was published on 2006 and it was entitled Data Mining in E-Learning (Romero and Ventura, 2006). Since then, an increasing number of books have been published (see Table 2), especially in the last years. We can see that during the first years (from 2006 to 2014) the terms Data Mining in Education and Educational Data Mining were used in the titles. Next (from 2015 to 2017) the terms Learning Analytics, Data Science, Big Data and Data Mining were also used. And in the last years the terms Learning Analytics is the most used in the titles. From all of them, the two most important books in the area are the Handbook of Educational Data Mining (Romero et al., 2010b) and the Handbook of Learning Analytics (Lang et al., 2017). Additionally, we want to highlight that there is also an online [1] Massive Online Open Textbook (MOOT) that contains all the resources as seen on Coursera (2013) and EdX (2015) Big Data and Education courses (Baker, 2015).

| Cover | Title | Authors | Year | Editorial | Pages |
|---|---|---|---|---|---|
| 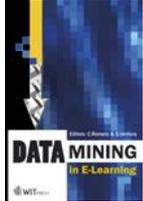 | Data Mining in Education | C. Romero & S. Ventura | 2006 | Wit Press | 299 |
| 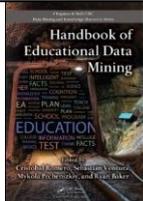 | Handbook of Educational Data Mining | C. Romero, S. Ventura., M. Pechenizky, R. Baker | 2010 | CRC Press, Taylor & Francis Group | 535 |
| 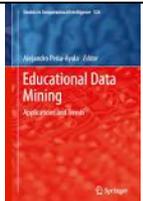 | Education Data Mining: Applications and Trends | A. Peña-Ayala | 2014 | Springer | 468 |
| 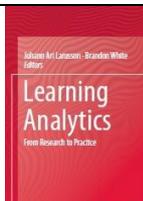 | Learning Analytics: From research to practice. | J.A. Larusson, B. White | 2014 | Springer | 195 |
| 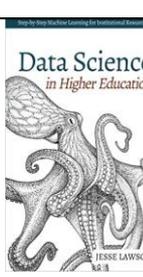 | Data Science in Higher Education: A Step-by-Step Introduction to Machine Learning for Institutional Researchers | J. Lawson | 2015 | CreateSpace Independent Publishing Platform | 226 |
| 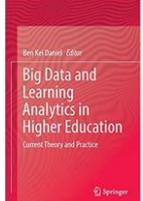 | Big Data and Learning Analytics in Higher Education: Current Theory and Practice | B.k. Daniel | 2016 | Springer | 272 |
| 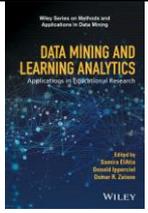 | Data Mining and Learning Analytics: Applications in Educational Research | S. ElAtia, D. Ipperciel, O.R. Zaïane | 2016 | Wiley | 320 |
| 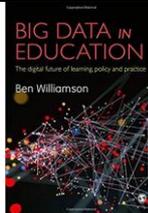 | Big Data in Education: The digital future of learning, policy and practice | B. Williamson | 2017 | SAGE Publications | 256 |
| 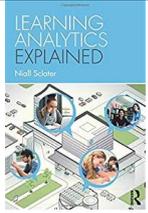 | Learning Analytics Explained | Niall Sclater | 2017 | Routledge | 290 |
| 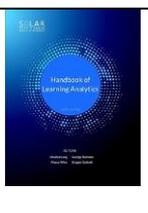 | Handbook of Learning Analytics | C. Lang, G. Siemens, A. Wise, D. Gašević | 2017 | SOLAR | 356 |
| 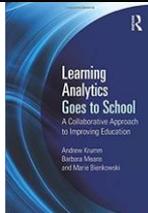 | Learning Analytics Goes to School | A. Krumm , B. Means, M. Bienkowski | 2018 | Routledge | 190 |
| 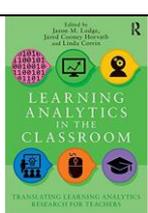 | Learning Analytics in the Classroom | J. Horvath, J. Lodge, L. Corrin | 2018 | Routledge | 314 |
| 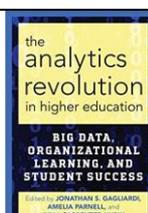 | The Analytics Revolution in Higher Education: Big Data, Organizational Learning, and Student Success | J. S. Gagliardi, A. Parnell, J. Carpenter-Hubin, R. L. Swing | 2018 | Stylus Publishing | 252 |
| 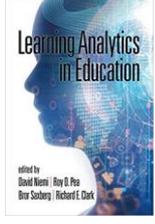 | Learning Analytics in Education | D. Niemi, R. D. Pea, B. Saxberg, R. E. Clark | 2018 | Information Age Publishing | 268 |
| 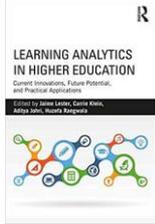 | Learning Analytics in Higher Education: Current Innovations, Future Potential, and Practical Applications | J. Lester, C. Klein, A. Johri, H. Rangwala | 2018 | Routledge | 216 |
| 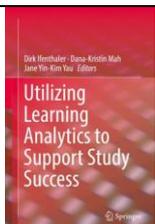 | Utilizing Learning Analytics to Support Study Success | D. Ifenthaler, D. Mah, Y. J. Yau | 2019 | Springer | 328 |
| 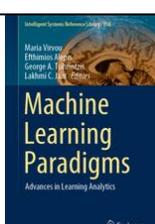 | Machine Learning Paradigms: Advances in Learning Analytics | M. Virvou, E. Alepis, G.A. Tsihrintzis, L.C. Jain | 2019 | Springer | 223 |
| 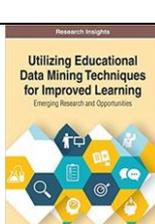 | Utilizing Educational Data Mining Techniques for Improved Learning: Emerging Research and Opportunities | C. Bhatt, P.S. Sajja, S. Liyanage | 2019 | IGI Global | 188 |

Table 2. Published books about EDM/LA.

There are several international and prestigious journals in which most of the papers about EDM and LA have been published (see Table 3). Of all of them, the two most specific journals are the Journal of Educational Data Mining[2] which was launched in 2009 and the Journal of Learning Analytics[3] which was launched in 2014.  We also want to notice that there are some just born new related journals such as: International Journal of Learning Analytics and Artificial Intelligence for Education (iJAI)[4] and Computer-Based Learning in Context (CBLC)[5] both launched in 2019.

---

[2] http://www.educationaldatamining.org/JEDM/
[3] https://solaresearch.org/stay-informed/journal/
[4] https://online-journals.org/index.php/i-jai
[5] https://solaresearch.org/stay-informed/journal/

| Journal Title | Number of papers | Impact Factor 2018 | Free and Open Access |
|---|---|---|---|
| Journal of Learning Analytics | 143 | - | Yes |
| Computers and Education | 81 | 5.627* | No |
| British Journal of Educational Technology | 65 | 2.588 ** | No |
| Journal of Educational Data Mining | 48 | - | Yes |
| Journal of Artificial Intelligence in Education | 47 | - | No |
| IEEE Transactions on Learning Technologies | 33 | 2.315* | No |
| Journal of Computer Assisted Learning | 32 | 2.451 ** | No |
| International Journal on Technology Enhanced Learning | 31 | - | No |
| User Modeling and User-Adapted Interaction | 27 | 3.400* | No |
| Internet and Higher Education | 26 | 5.284** | No |
| Computer Applications in Engineering Education | 26 | 1.435 * | No |

Table 3. Top related journals about EDM/LA (*JCR Science Edition, **JCR Social Science Edition, 01-01-2019).

Finally, in order to show the grow interest in EDM and LA during the last two decades, Figure 2 shows the number of papers or results that return a freely accessible web search engine such as *Google Schoolar* when searching the exact term "Educational Data Mining" or "Learning Analytics" in each year from 2000 to 2018. As can be seen, both numbers grow in an exponential way, showing the high interest in both topics. And although EDM has more references than LA until 2011, then LA surpass EDM. This fact can be explained by the temporal distribution of the next important events in EDM and LA history such as: the first Workshop about EDM in the Association for the Advancement of Artificial Intelligence 2005 Conference,  the first book related with EDM/LA (Romero and Ventura, 2006), the first Educational Data Mining Conference in  Montreal (Canada) 2009, the handbook of EDM (Romero et al., 2010b),  the first Learning Analytics&Knowledge in Banf (Canada) 2011, the first LA Summer Institute in Palo Alto (USA) 2013,  the first Learning at Scale Conference in Atlanta (USA) 2015, and the handbook of LA (Lang et al., 2017). This greater increment in the number of papers that use the term "Learning Analytics" rather than "Educational Data Mining" produce that a bibliometric approach (Dormezil et al. 2019) conclude it is more accurate to describe what appears to be two domains (i.e. Educational Data Mining and Learning Analytics) as one domain (i.e. Learning Analytics) with  one prominent subset (i.e. Educational Data Mining).

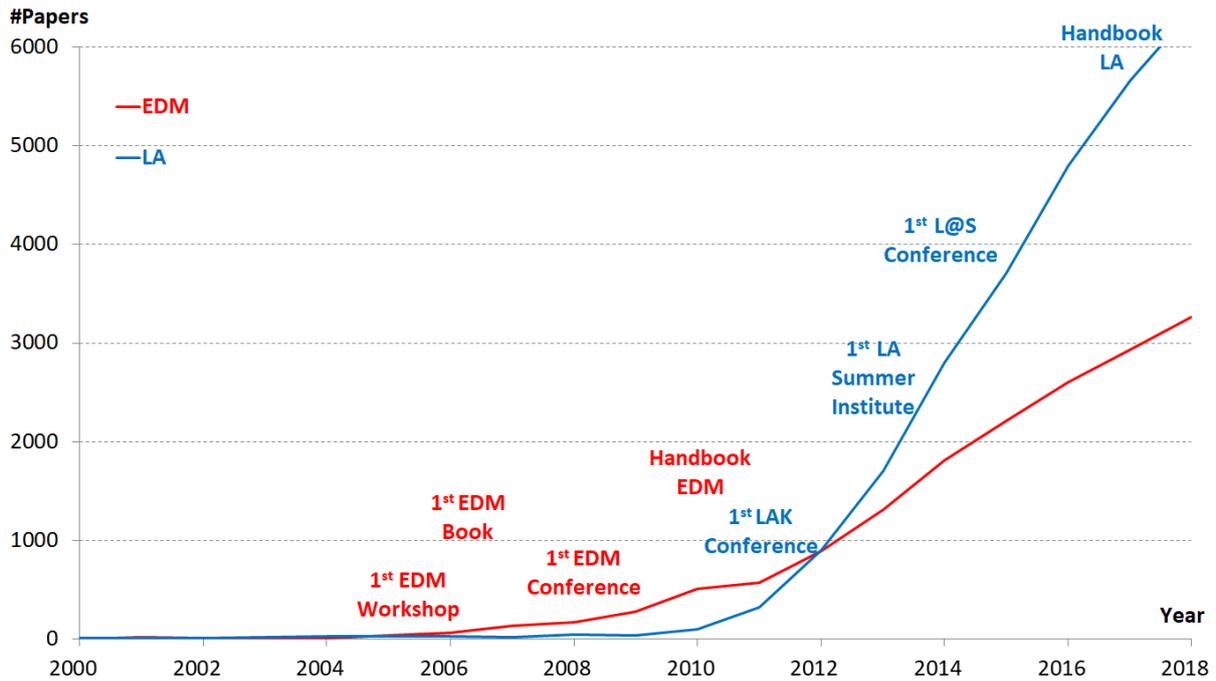

Figure 2. Number of papers and main events about EDM/LA terms in Google Schoolar by year (01-01-2019).

Finally, the most cited papers in EDM and LA are shown in Table 4. We can see that four in ten papers are reviews/surveys. Analyzing this important type of papers, the first and most popular review of EDM research was published by Romero and Ventura (2007), and it was followed by a more complete (Romero and Ventura, 2010a) and a more comprehensible (Romero and Ventura, 2013) reviews by the same authors. Another popular review was presented on the inaugural issue of EDM journal (Baker and Yacef, 209). An important report was published by the U.S. Office of Educational Technology about how to enhance teaching and learning through EDM and LA (Bienkowski et al., 2012). The differences between EDM and LA are deal in other highly cited review (Siemens and Baker, 2012). And finally, two good specific reviews about LA provide us an introductory to this area (Ferguson, 2012) and an analysis of the citation networks of the area (Dawson et al. 2014).

| Paper Title | Reference | Num. cites * | Num cites ** |
|---|---|---|---|
| Educational data mining: A survey from 1995 to 2005 | (Romero and Ventura 2007) | 1489 | 662 |
| Educational data mining: a review of the state of the art | (Romero and Ventura 2010a) | 1367 | 631 |
| The state of educational data mining in 2009: A review and future visions | (Baker and Yacef 2009) | 1199 | - |
| Penetrating the fog: Analytics in learning and education | (Siemens and Long, 2011) | 1138 | - |

| Data mining in course management systems: Moodle case study and tutorial | (Romero et al., 2008) | 1105 | 470 |
| Learning analytics: drivers, developments and challenges | (Ferguson, 2012) | 691 | 328 |
| Learning analytics and educational data mining: towards communication and collaboration | (Siemens and Baker 2012) | 589 | 224 |
| Course signals at Purdue: Using learning analytics to increase student success | (Arnold and Pistilli, 2012) | 569 | 206 |
| Translating learning into numbers: A generic framework for learning analytics | (Greller and Drachsler, 2012) | 547 | 221 |
| Mining educational data to analyze students' performance | (Baradwaj and Pal, 2012) | 543 | - |

Table 4. Top-10 most cited papers about EDM and LA  (*Google Schoolar, **SciVerse Scopus, 01-01-2019).

## EDM/LA KNOWLEDGE DISCOVERY CYCLE

The process of applying EDM/LA is a cycle application  of the general Knowledge Discovery and Data Mining (KDD) process (see Figure 3) although there are some important differences with specific characteristics in each step as is described in the subsections below.

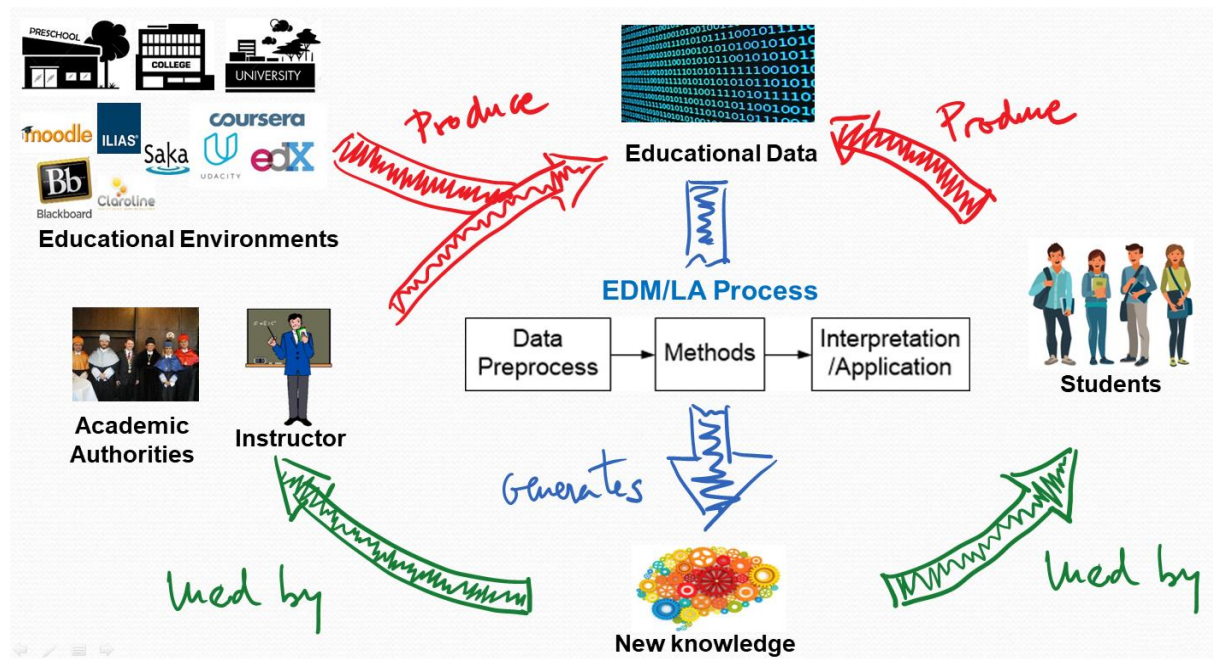

Figure 3. EDM/LA knowledge discovery cycle process.

**Educational Environment**

Depending on the type of the educational environment (traditional classroom education, computer-based or blended learning education) and the information system used, such as: LMS (Learning Management System), ITS (Intelligent Tutoring System), MOOC (Massive Open Online Course), etc., different kinds of data can be collected in order to resolve different educational problems (Romero et al., 2011).

**Educational Data**

Educational data are gathered (Romero et al. 2014) from different sources such as the interaction between instructors, students and the educational (e.g., navigation behavior, input in quizzes, interactive exercises, forum messages, etc.) administrative data (e.g., school and teacher information), demographic data (e.g., gender, age, etc.), student affectivity (e.g., motivation, emotional states), etc. Educational environments can store a huge amount of potential data from multiple sources with different formats and with different granularity levels (from coarse to fine grain) or multiple levels of meaningful hierarchy (keystroke level, answer level, session level, student level, classroom level, and school level) that provide more or less data (see Figure 4). Gathering and integrating all this raw data for mining are non-trivial tasks on their own and thus a preprocessing step is necessary.

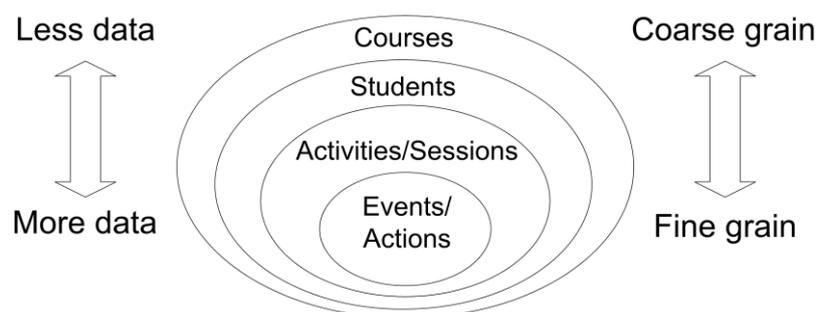

Figure 4. Different levels of granularity and their relationship to the amount of data.

**Preprocessing**

Data pre-processing is a hard and complicated task, and sometimes the data pre-processing itself takes up more than half of the total time spent solving the data mining problem (Bienkowski et al. 2012). Educational data available (raw, original or primary data) to solve a problem normally are not in the appropriate form (or abstraction). And so, it is necessary to convert the data to an appropriate form (modified data) for solving each specific educational problem. This includes choosing what data to collect, focusing on the questions to be answered, and making sure the data align with the questions. Traditional preprocessing tasks are applied to educational data with some specific issues (Romero et al., 2014) such as the next ones. Feature engineering for generating and selecting attributes/variables with information about the students is very important. Normally we can reduce and transform all available attributes into a summary table for better analysis. Continuous attributes are normally transformed/discretized into categorical attributes in order to improve their comprehensibility. Finally, it is important to maintain and protect the confidentiality of student by anonymizing data and deleting all personal information (not useful for mining purposes) such as name, e-mail, telephone number, etc. In this line, we can have into consideration use guidelines about ethical issues, data privacy, informed consent, etc. when using educational data (Pardo and Siemens, 2014).

**Methods and techniques**

The majority of traditional Data Mining techniques including but not limited to visualization, classification, clustering, and association analysis techniques have been already applied successfully in the educational domain (Baker, 2010). Nevertheless, educational systems have also some special characteristics (hierarchical and longitudinal data) that require a specific treatment of the mining problem and preprocess of the data. There are a wide range of EDM and LA methods and techniques used for solving different educational problems as described in the Methods section below.

**Interpretation and application of the new knowledge**

Taking action is the ultimate goal of any learning analytics process and the results of follow-up actions will determine the success or failure of our analytical efforts (Siemens, 2013). So, the discovered new knowledge by the EDM/LA methods have to be used by instructors and academic authorities to make interventions and decisions in order to improve student learning performance. It is very important that the previous models obtained by the EDM/LA process were comprehensible in order to be useful for the decision-making process. In this line, white-box DM models such as decision trees are preferable to black-box models such as neural networks as they are more accurate but less comprehensible. Visualization techniques are also very useful for showing results in a way that is easier to interpret. Recommender systems are very useful for providing explanations and recommendations both to students and a non-expert user in EDM/LA such as instructors.

## EDUCATIONAL ENVIRONMENTS AND DATA

There is a wide variety of educational environments (see Figure 5) such as traditional education, computer-based education and blended learning. Each one of them provides different data sources (Romero and Ventura, 2007).

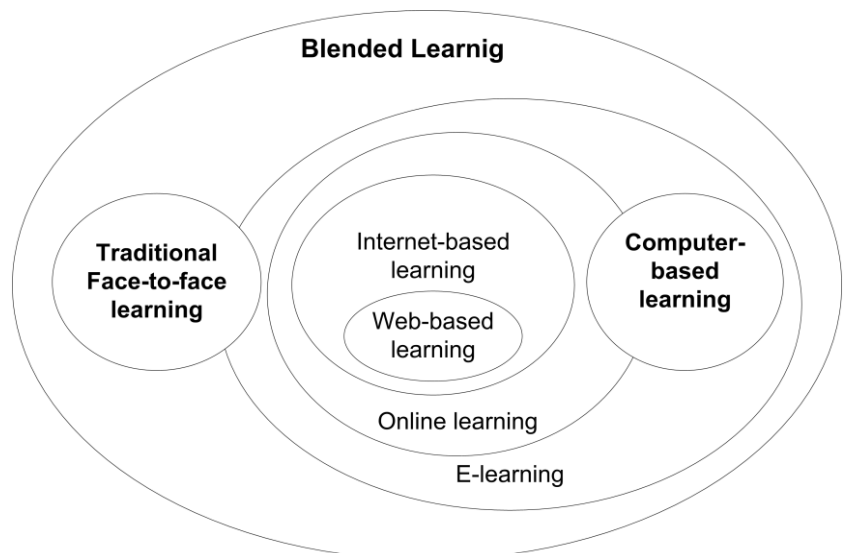

Figure 5. Types of educational environments and systems.

**Traditional face-to-face education**

Traditional education or back-to-basics are the most widely-used educational system, based mainly on face-to-face contact between educators and students organized through lectures, class discussion, small groups, individual seat work, etc. Traditional education systems are classified in different levels by UNESCO[6] as we can see in Table 5. These systems gather information on student attendance, marks, curriculum goals, class, schedule information, etc. Finally, it is important to note that all these traditional systems can also use computer-based educational systems as a complementary tool to face-to-face sessions.

| Level | Principal characteristics |
|---|---|
| Pre-primary education | Designed for children from age 3 to the start of primary education. |
| Primary education or first stage of basic education | Normally starting between the ages of 5 – 7. |
| Lower secondary education or second stage of basic education | Designed for children from ages 8-14 to complete basic education. |
| Upper secondary education | More specialized education beginning at age 15 or 16 years. |
| Post-secondary non-tertiary education | Programmes that straddle the boundary between upper- and post-secondary education from an international point of view. |

| First stage of tertiary education | Tertiary programmes having an educational content more advanced than those offered at previous levels. |
| Second stage of tertiary education | Tertiary programmes leading to the award of an advanced research qualification, e.g. Ph.D. |

Table 5. International levels of education.

**Computer-based educational systems**

Computer Based Education (CBE) means using computers in education to provide direction, to instruct or to manage instructions given to the student. CBE systems were originally simple stand-alone educational applications that ran on a local computer. But the global use of Internet and the application of Artificial Intelligence (AI) techniques have led to today's plethora of new web-based intelligent educational systems. Some examples of computer-based educational systems used currently are listed and described in Table 6.

| System | Description |
|---|---|
| Adaptive and Intelligent Hypermedia System (AIHS) | These systems attempt to be more adaptive by building a model of the goals, preferences, and knowledge of each individual student and using this model throughout interaction with the student in order to adapt to the needs of that student. The data recorded by AIHs are similar to ITS data. |
| Intelligent Tutoring System (ITS) | ITSs provide direct customized instruction or feedback to students by modeling student behavior and changing its mode of interaction with each student based on its individual model. Normally, it consists of a domain model, student model and pedagogical model. ITSs record all student-tutor interaction (mouse clicks, typing, and speech). |
| Learning Management System (LMS) | Suites of software that provide course-delivery functions: administration, documentation, tracking, and reporting of training programs, classroom and online events, e-learning programs, and training content. They record any student activities involved, such as reading, writing, taking tests, performing tasks in real, and commenting on events with peers. |
| Massive Open Online Course (MOOC) | It refers to a web-based class designed to support a large number of participants. It can deliver learning content online to any person who wants to take a course, with no limit on attendance. They store the same information that LMS. |
| Test and quiz system | The main goal of these systems is to measure the students' level of knowledge with respect to one or more concepts or subjects by using a series of questions/items and other prompts for the purpose of gathering information from respondents. They store a great deal of information about students' answers, calculated scores, and statistics. |
| Other types | Wearable learning systems, learning object repositories, concept maps, social networks, WIKIs, forums, educational and serious games, virtual and augmented reality systems, etc. They store different type of information about the interaction with the students. |

Table 6. Examples of computer-based educational systems.

**Blended Learning Systems**

Blended Learning (BL) environments combine face-to-face instruction with computer-mediated instruction. The terms "blended learning", "hybrid learning" and "mixed-mode instruction" are often used interchangeably in research literature. Blended courses increased access, convenience and it provides more flexibility and freedom compared to face to face courses by transforming significant amount of face to face sessions online. These systems gather information about the two previous face-to-face and the computer-based systems.

## TOOLS AND DATASETS

Nowadays, there is a wide array of well-known general purpose tools and frameworks that can be used for the purposes of conducting EDM and LA research (Slater at al., 2017) such as: Rapidminer, Weka, SPSS, Knime, Orange, Spark Lib, etc. However, these tools aren't easy for educators to use due to they are required to select the specific method/algorithm to apply/use and to provide the appropriate parameters in advance in order to obtain good results/models. So, the educators must possess a certain amount of expertise in order to find the right settings (Romero and Ventura, 2013). A solution to this problem is the use of some of the specific EDM/LA available software tools (see Table 7). However, they only work with specific data in order to solve specific educational problems.

| Name | URL | Description |
|------|-----|-------------|
| DataShop | https://pslcdatashop.web.cmu.edu/ | It provides both a central repository to secure and store research data, and a set of analysis and reporting tools. |
| GISMO | http://gismo.sourceforge.net/ | Graphical interactive monitoring tool that provides useful visualization of students' activities in online courses to instructors. |
| Inspire | https://moodle.org/plugins/tool_inspire | Moodle Analytics API that provides descriptive and predictive analytics engine, implementing machine learning backends. |
| LOCO-Analyst | http://jelenajovanovic.net/LOCO-Analyst/ | Tool aimed at providing teachers with feedback on the relevant aspects of the learning process taking place in a web-based learning environment. |
| Meerkat-ED | http://www.reirab.com/MeerkatED | Tool for analyzing students' activity in a course offered on computer-supported collaborative learning tools. |
| MDM Tool | http://www.uco.es/kdis/research/software/ | Framework for apply some data mining techniques in Moodle 2.7 version. |
| Performance Plus | https://www.d2l.com/higher-education/products/performance/ | Package for delivering powerful analytics tools to help administrators, educators, and learners save quality time while maximizing |

| | | impact and driving success. |
|---|---|---|
| SNAPP | https://web.archive.org/web/20120321212021/http://research.uow.edu.au/learningnetworks/seeing/snapp/index.html | Tool that allows users to visualize the network of interactions resulting from discussion forum posts and replies. |
| Solutionpath StREAM | https://www.solutionpath.co.uk/ | Real-time system that leverage predictive models to determine all facets of student engagement |

Table 7. Examples of specific EDM/LA tools.

Most of the EDM/LA researchers normally use their own data for solving their specific educational problems. But it is a hard and very time consuming task to gather and preprocess educational data (Romero et al., 2014). So, another option is to use some of the public datasets that are currently available for free download in Internet as we show in Table 8.

| Datasets | URL | Description |
|---|---|---|
| ASSISTments Competition Dataset | https://sites.google.com/view/assistmentsdatamining/home | Competition where data miners can try to predict an important longitudinal outcome using real-world educational data. |
| Canvas Network dataset | https://dataverse.harvard.edu/dataset.xhtml?persistentId=doi:10.7910/DVN/1XORAL | De-identified data from Canvas Network open courses (running January 2014 - September 2015), along with related documentation. |
| DataShop | https://pslcdatashop.web.cmu.edu/index.jsp?datasets=public | LearnSphere's DataShop provides a central repository to secure and store research ITS data and set of analysis and reporting tools. |
| Educational Process Mining Dataset | https://archive.ics.uci.edu/ml/datasets/Educational+Process+Mining+(EPM)%3A+A+Learning+Analytics+Data+Set | Students' logs during sessions over a simulation environment in digital electronics. |
| HarvardX-MITx dataset | https://dataverse.harvard.edu/dataset.xhtml?persistentId=doi:10.7910/DVN/26147 | De-identified data from the first year of MITx and HarvardX MOOC courses on the edX platform along with related documentation. |
| KDD Cup 2010 Dataset | https://pslcdatashop.web.cmu.edu/KDDCup/ | Challenge to predict student performance on mathematical |

| | | problems from logs of student interaction with ITS. |
|---|---|---|
| Learn Moodle dataset | https://research.moodle.net/158/ | Anonymized data from the "Teaching with Moodle August 2016" course from learn.moodle.net. |
| MOOC-Ed Dataset | https://dataverse.harvard.edu/datas et.xhtml?persistentId=doi:10.7910/D VN/ZZH3UB | Communications taking place between learners in two offerings of the Massively Open Online Course for Educators (MOOC-Eds). |
| NAEP Data Mining Competition 2019 | https://sites.google.com/view/data miningcompetition2019/dataset | Competition for measuring students' test taking activities, and helps develop and test evaluation methods for educational analysis. |
| NUS Multi-Sensor Presentation Dataset | http://mmas.comp.nus.edu.sg/NUS MSP.html | It contains real-world presentations recorded in a multi-sensor environment. |
| Open University Learning Analytics Dataset | https://analyse.kmi.open.ac.uk/open _dataset | It contains data about courses, students and their interactions with Moodle for seven selected courses. |
| Student Performance Dataset | https://archive.ics.uci.edu/ml/datas ets/Student+Performance | This data approach student achievement in secondary education of two Portuguese schools. |
| xAPI-Educational Mining Dataset | https://www.kaggle.com/aljarah/xA PI-Edu-Data | Students' Academic Performance Dataset collected from e-learning system called Kalboard 360. |

Table 8. EDM/LA public datasets.

We want to highlight DataShop (Koedinger et al. 2010) as one of the first and biggest dataset that also provides a tool for researching about ITS. As we see in Table 8, currently there are not many public datasets available and they are not from all the types of educational environments (most of them are from e-learning systems). So, we think that in the future it will be very useful to develop a specific EDM/LA datasets repository similar to the general UCI Machine Learning Repository[7]. It is important to note that these public datasets must be portable and they must consider principles of data ethics, privacy, protection and consent (Fergunson et al. 2016). The idea of data portability is that educational institution/instructors/researchers should not have their own data stored in "silos" or "walled gardens" that are incompatible with one another but to use standards such as Experience API (xAPI) [8]or IMS Caliper[9].

## METHODS AND APPLICATIONS

There is a wide range of popular methods (See Table 9) within EMD and LA (Romero and Ventura, 2013) (Baker and Inventado, 2014) (Bakhshinategh et al., 2018) for solving educational problems or application. Most of these techniques are widely acknowledged to be universal across types of data mining, such as visualization, prediction, clustering, outlier detection, relationship mining, causal mining, social network analysis, process mining and text mining. And others have more prominence within education, such as the distillation of data for human judgment, discovery with models, knowledge tracing and non-negative matrix factorization.

| Method | Goal/description | Key applications |
|---|---|---|
| Causal Mining | To find causal relationship or to identify causal effect in data. | Finding what features of students' behavior cause learning, academic failure, drop out, etc. |
| Clustering | To identify groups of similar observations. | Grouping similar materials or students based on their learning and interaction patterns. |
| Discovery with models | To employ a previously validated model of a phenomenon as a component in another analysis. | Identification of relationships among student behaviours and characteristics or contextual variables. Integration of psychometric modelling frameworks into machine-learning models. |
| Distillation of data for human judgment | To represent data in intelligible ways using summarization, visualization and interactive interfaces. | Helping instructors to visualize and analyze the ongoing activities of the students and the use of information. |
| Knowledge tracing | To estimate student mastery of skills, employing both a cognitive model that maps a problem-solving item to the skills required, and logs of students' correct and incorrect answers as evidence of their knowledge on a particular skill. | Monitoring student knowledge over time. |

| Method | Goal/description | Key applications |
|---|---|---|
| Nonnegative matrix factorization | To define a matrix of positive numbers with student test outcome data that may be decomposed into a matrix of items and a matrix of student mastery of skills. | Assessment of student skills. |
| Outlier detection | To point out significantly different individuals. | Detection of students with difficulties or irregular learning processes. |
| Prediction | To infer a target variable from some combination of other variables. Classification, regression and density estimation are types of prediction methods. | Predicting student performance and detecting student behaviours. |
| Process mining | To obtain knowledge of the process from event logs. | Reflecting students' behaviour based on traces of their evolution through the educational system. |
| Recommendation | To predict the rating or preference a user would give to an item. | To make recommendations to students with respect to their activities or tasks, links to visits, problems or courses to be done, etc. |
| Relationship mining | To study relationships among variables and to encode rules. Association rule mining, sequential pattern mining, correlation mining and causal data mining are the main types. | Identifying relationships in learner behaviour patterns and diagnosing student difficulties. |
| Statistics | To calculate descriptive and inferential statistics. | Analyzing, interpreting and drawing conclusions from educational data. |
| Social network analysis | To analyze the social relationships between entities in networked information. | Interpretation of the structure and relations in collaborative activities and interactions with communication tools. |
| Text mining | To extract high-quality information from text. | Analysing the contents of forums, chats, web pages and documents. |
| Visualization | To show a graphical representations of data. | To produce data visualizations that help communicate results of |

| Method | Goal/description | Key applications |
|--------|------------------|------------------|
| | | EDM/LA research to educators. |
| Nonnegative matrix factorization | To define a matrix M of positive numbers with student test outcome data that may be decomposed into two matrices: Q, which represents a matrix of items, and S, which represents student mastery of skills. | Assessment of student skills. |

Table 9: Most popular EDM/LA methods.

However, the number of possible objectives or educational problems in EDM/LA is huge and this taxonomy does not cover all the possible tasks. In fact, there are many more specific objectives depending on the viewpoint of the final user. Although an initial consideration seems to involve only two main groups of potential users/stakeholders – the learners and the instructors – there are actually more groups involved with many more objectives, as can be seen in Table 10. And, in order to show more examples of the most promising applications of the EDM/LA, Table 11 shows some of the current hot topics or more interesting problems in the area.

| User/Stakeholders | Objectives |
|-------------------|-----------|
| Learners or Students | Learners are interested in understanding student needs and methods to improve the learner's experience and performance. |
| Educators or Instructors | Educators attempt to understand the learning process and the methods they can use to improve their teaching methods. |
| Scientific Researchers | Researchers focus on the development and the evaluation of EDM/LA techniques for effectiveness. |
| Administrators or Academic Authorities | Administrators are responsible for allocating the resources for implementation in institutions. |

Table 10. Example of users/stakeholders and objectives.

| Topics of Interest | Description | Reference |
|--------------------|-------------|-----------|
| Analyzing educational theories | To analyze how learning theories and learning analytics could be integrated in educational research. | (Wong et al., 2019) |
| Analyzing pedagogical strategies | To analyze and explore the application and effect of pedagogical strategies with EDM/LA techniques. | (Shen et al. 2018) |
| Analyzing programming code | To apply EDM/LA techniques focused on analysing code from programming courses, programming assignments/submissions, etc. | (Li and Edwards, 2018) |
| Collaborative | To analyze collaborative learning and to predict the team | (Hernández-García |

| learning and teamwork group | grade in teamwork groups. | et al. 2018) |
|---|---|---|
| Curriculum Mining/Analytics | To analyze program structure, course grading and administrative curricular data in order to improve curriculum development, program quality, etc. | (Hilliger et al,. 2019) |
| DashBoards and visual learning analytics | To apply a visualization techniques to explore and understand relevant user traces that are collected in (online) environments and to improve (human) learning. | (Millecamp, 2019) |
| Deep Learning | To apply neural network architectures with multiple layers of processing units in EDM/LA research area. | (Hernández-Blanco, 2019) |
| Discovery causal relationships | To find causality relationship among attributes in an educational data set. | (de Carvalho et al., 2019) |
| Early Warning Systems | To predicting student's performance and students at risk as soon as possible in order to intervene early to facilitate student success. | (Cano and Leonard, 2019) |
| Emotional Learning Analytics | To study affect during learning and the importance of emotion to learning. | (D'Mello et al., 2017) |
| Evaluating the efficacy of interventions | To evaluate the efficacy of interventions, data-driven student feedback, actionable advice, etc. | (Sonderlund et al, 2018) |
| Feature Engineering Methods | To build automatically attributes or students features using machine learning techniques. | (Botelho et al. 2019) |
| Game Learning Analytics | To apply data-mining and visualization techniques to player interactions in serious games. | (Alonso-Fernández, 2019) |
| Interpretable and explanatory learner models | To develop "white box" interpretable, explanatory, usable, and highly comprehensible leaner models. | (Rose et al. 2019) |
| Learning foreign language | To apply EDM/LA techniques for improving of foreign language learning. | (Bravo-Agapito et al. 2019) |
| Measuring Self-Regulated Learning | To apply EDM/LA techniques to measure self-regulated learning feature and behaviors in students. | (ElSayed et al., 2019) |
| Multimodal Learning | To apply of machine learning and increasingly affordable sensor technologies for providing new types learning | (Spikol et al., 2017) |

| Analytics | insights that happen across multiple contexts. | |
|---|---|---|
| Orchestrating learning analytics | To study the adoption, implications for practice and other factors in ongoing LA adoption processes at classroom level. | (Prieto et al. 2019) |
| Providing personalized feedback | To generate personalized feedback automatically or semi-automatically to support the student learning. | (Pardo et al. 2019) |
| Sentiment Discovery | To automatically identify the underlying attitudes, sentiments, and subjectivity in learners and learning resources. | (Han et al. 2019) |
| Transfer Learning | To develop models that can be transferable or applied to other similar courses/institutions/etc. | (Ding et al. 2019a) |
| Understanding navigation paths | To discover process-related knowledge and navigational learning from event logs recorded by e-learning systems. | (Bogarin et al., 2018) |
| Writing analytics | To apply text mining and analytics tools to text data from forums, chats, social networks, assessments, essays, etc. | (Ferreira-Mello et al., 2019) |

Table 11. Some current applications or topics of interest of EDM/LA research community.

## CONCLUSIONS AND FUTURE TRENDS

EDM and LA are two inter-disciplinary communities of computer scientists, learning scientists, psychometricians, and researchers from other areas with the same objective of improve learning starting from data. This area has grown quickly in the last two decades with two different annual conferences (EDM and LAK), two specific journals (JEDM and JLA), and increasing number of books, papers and surveys/reviews. Additionally, they are a current move from the lab to the general market for using EDM/LA by educational institutions and schools worldwide and it is expected that in 2020 all education research involves analytics and data mining (Ryan and Inventado, 2014). All this indicates us that EDM/LA will become soon a mature area that will be widely used not only by researchers but also by instructors, educational administrators and related business from all over the world. EDM/LA has impacted our understanding of learning and produced insights that have been translated to mainstream practice or contributed to theory. The research in this area has developed in the study focus and sophistication of analyses, but the impact on practice, theory and frameworks have been more limited (Dawson et al., 2019). This is due to it is necessary the stimulus from existing research organizations (such as SoLAR and IEDM), funding agencies, collaborations and the active promotion of established works such as the LACE initiative[10] in order to increase the impact on practice and to move from exploratory models to more holistic and integrative systems-level research.

---

[10] http://www.laceproject.eu/lace/

As for future trends in the area, our previous survey (Romero and Ventura, 2013) proposed two trends. However, one of them has not been completely achieved yet and the other remains a challenge.  The first one was that it is necessary more freely available EDM tools in order to a wider and broader population can use them. As we can see in section Tools and Datasets, currently there is now a wide array of specific purpose tools. However, it still needs to develop general purpose EDM/LA tools to apply several tasks for solving different educational problems from the same interface/tool. And it is also necessary to improve the portability of the obtained models from these tools. The second one was that educators and institutions should develop a data-driven culture of using data for making instructional decisions and improving instruction. However, according to a recent report (Joksimović et al. 2019), the majority of the institutions continue aware of the benefits provided by the analysis of large-scale data about student learning. In order to address the complexity of scaling learning analytics Dawson and colleagues (Dawson et al., 2018) argued for the inclusion of new forms of leadership models in education to stimulate and promulgate systemic change. And specific Learning Analytics challenges to overcome in higher education institutions are proposed and grouped into the next seven categories (Leitner et al. 2019):

1. Purpose and gain. It is necessary to make the goals of the LA initiative transparent, clarifying exactly what is going to happen with the information and explicitly what is not.

2. Representation and actions. To choose the right environment for the learner's feedback, the correct visualization technique to provide recommendations and results to the students.

3. Data. A policy needs to be created for LA that aligns with the organization's core principles. Transparent communication about where the data are stored, what is being done to ensure data security and privacy and how the data are evaluated and used (and by whom) is essential.

4. IT infrastructure. Efforts should be made from the beginning to search for possible solutions to set up the necessary internal or external IT infrastructure and contact and establish connections with relevant people.

5. Development and operation. Scalability is maybe one of the most frequently underestimated problems in today's IT industry. A distinction must be made as to whether processes have to be carried out manually, semi-automatically or fully automatically.

6. Privacy. All LA implementations have to ensure the privacy of the involved parties. The general lifetime of personal data is a topic that requires further discussion.

7. Ethics.  LA implementers must find a suitable way to meet high ethical standards and ensure a beneficial outcome for all stakeholders.

Additionally, the Baker Learning Analytics Prizes (BLAP) proposes the next six specific research problems as challenges of EDM/LA area (Baker, 2019):

1. Transferability: The (learning system) Wall Transfer student model from learning system A to learning system B. Improve an already-good student model in learning system B. Change behavior of learning system B in runnable fashion.

2. Effectiveness: Differentiating Interventions and Changing Lives Publicize criterion for intervention. Assign students to control or experimental group. Use analytics, only within experimental group, to assign intervention Collect longer-term outcome measure. Demonstrate that experimental/analytics-intervention group performs better than experimental/analytics-no-intervention group. But that experimental/analytics-no-intervention group does not perform better than control/analytics-no-intervention group.

3. Interpretability: Instructors Speak Spanish, Algorithms Speak Swahili. Take a complex model of a learning analytics phenomenon. Develop a no-human-in-the-loop method of explaining the model. Present the explanation to five (new) data scientists and users. Ask the participants to explain what decision the model will make, and why, for five case studies. Code the explanations of the model's decisions. Verify if the data scientists and users interpret the model the same way for the case studies.

4. Applicability: Knowledge Tracing Beyond the Screen. Take data from at least four students completing learning activity together. Model at least four distinct skills for each student. Predict immediate future performance for these skills.

5. Generalizability: The General-Purpose Boredom Detector. Build an automated detector of affect. Demonstrate that the detector works for an entirely new learning system with different interactions and with AUC ROC >= 0.65.

6. Generalizability: The New York City and Marfa Problem. Build an automated detector for a commonly-seen outcome or measure. Collect a new population distinct from the original population. Demonstrate that the detector works for the new population with degradation of quality under 0.1 (AUC ROC, Pearson/Spearman correlation) and remaining better than chance.

Finally, we want to propose some personal visionary ideas that, in our opinion, might be very promising trends of EDM/LA in a near future:

- Taking into account all students' personal data through their whole life. Currently, information considered in EDM/LA is mainly based only in the interaction of students with a single educational environment, but in a near future thanks to big data and IoT (Al-Emran et al. 2020) we will be able to have information available for each student from their birth to this very moment and on real time. It will imply the integration of not only the traditional performance and usage data gathered from all the previous institutions and educational environments each student has used, but also the information about the personal status of each student from different points of view such as medical, familiar, economical, religious, sexual, relationship, emotional, psychological, etc. All these data could be gathered from multiple available sources and they could be fusioned (Ding et al. 2019b) in order to can be used for improving and personalizing the learning process to each single student in each specific moment of their live to a new level of precision.
- Applying and integrating EDM/LA to upcoming technological educational environments. Over the last decade the great advances on innovative technologies has enabled the development of new educational systems from mobile and ubiquitous to virtual reality, augmented reality environments and holograms (Cerezo et al., 2019). In the next few decades, quantum leaps will be associated to the application of artificial intelligence. In this context it is not wrong to think that instructors could be replaced by machines without students noticing the change thanks to current progresses in intelligent humanoids robots (Newton and Newton, 2019) and conversational agents or voice assistant interfaces in educational environments (Kloos et al., 2019). But these systems will need EDM/LA techniques for analyzing tons of data and generating portable analytics models[11] in real time in order to address the specific forthcoming educational challenges of these future technological environments.

- Analyzing and mining data directly gathered from students' brain for a better understanding of the learning. The brain is the key factor to really understand how students learn. The promising advances in human neuroscience and pervasive neurotechnology (brain-computer interfaces, BCI) are giving rise to unprecedented opportunities for getting, collecting, sharing and manipulating any kind of information gathered from the human brain (Williamson, 2019). In a near future, these intimate data about student's psychological state and neural activity could be analysed and mining in real-time thanks to upcoming small high-quality EEG (Electroencephalography) devices. These brain data, together with other multimodal data (Giannakos et al. 2019) could be integrated and used by EDM/LA researchers in order to can achieve a fully understanding of students' learning process.

## FUNDING INFORMATION

This research is supported by projects of the Ministerio de Ciencia e Innovación TIN2017-83445-P.

---

[11] http://dmg.org/